\documentclass[twocolumn,american,aps,showpacs]{revtex4}
\usepackage[T1]{fontenc}
\usepackage[latin1]{inputenc}
\usepackage{graphicx}

\makeatletter

\usepackage{graphicx}

\usepackage{babel}
\makeatother
\begin{document}

\title{The relation between nuclear charge radii and other parameters characterizing
the nuclear drops.}

\author{Zbigniew Sosin}

\email{ufsosin@cyf-kr.edu.pl}

\affiliation{Jagellonian University, M. Smoluchowski Institute of Physics, Reymonta
4, PL-30059 Krak\'{o}w, Poland }

\begin{abstract}
Correlation between the rms nucleus charge radius and their deformation,
isospin asymmetry, mass and charge numbers are presented. Four parameters
radius parametrization formula is proposed. Ratio of experimental
to theoretical value distribution exhibit variation equal to 0.0044,
i.e. it corresponds to the value of experimental errors. Observed
correlations can impose interesting constraints on the form of the
nuclear equation of state.
\end{abstract}


\maketitle
In the laboratory framework the nuclear matter is available only as
the atomic nuclei, which can be viewed as the liquid drops of nuclear
matter. Because the dimension and shape of nuclei reflect complicated
interactions inside nuclear matter it is of great importance to find
correlations between the parameters describing the nuclear drops.
In this letter we demonstrate new simple relation between these parameters,
based on extensive data for 574 heavy nuclei. In particular, we choose
the relation between nuclear charge radii and their deformation, isospin
$I$, mass $A$ and charge numbers $Z$. In order to describe the
nucleus deformation one can introduce different kinds of the shape
parametrization~\cite{Moller_95}. In our approach description of
deformations is limited to axially symmetric quadruply deformed shapes
and therefore for every nuclei the respective $\beta$ parameters
are attributed. Thess $\beta$ parameters are assumed to be taken
from model calculations presented in \cite{Moller_95}.

\begin{figure}[h]
\includegraphics[%
  width=8.5cm]{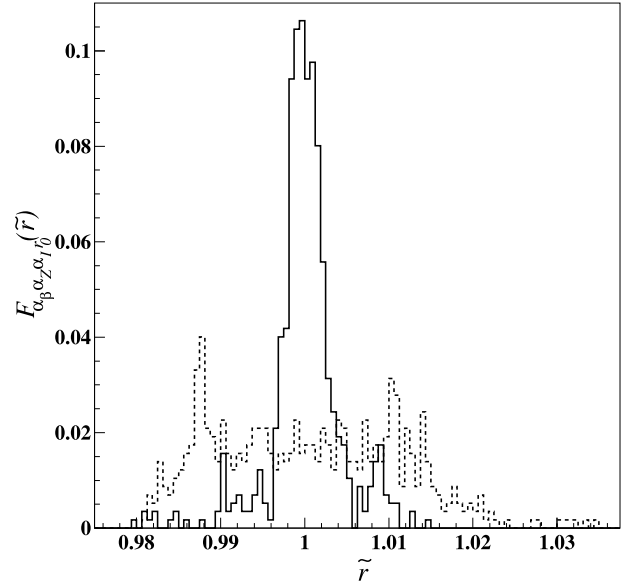}

\caption{\label{r1.epsi} The distribution $F_{\alpha_{\beta}\alpha_{Z}\alpha_{I}r_{0}}(\tilde{r})$.
The solid lines represents the distribution of $\tilde{r}$ for parameters
$\alpha_{\beta}$, $\alpha_{Z}$ ,$\alpha_{I}$, $r_{0}$ given by
(\ref{fit}). Dashed line represents this some distribution for parameters
$\alpha_{\beta}=0$, $\alpha_{Z}=0$, $\alpha_{I}=0$ and $r_{0}=0.945$
what for such case is the distribution of $r^{exp}/r_{0}$.}
\end{figure}

The reduced nuclear charged radius can be defined as\begin{equation}
r(Z,I,\beta)=\frac{\left\langle R^{2}\left(\beta Z,I\right)\right\rangle ^{1/2}}{A^{1/3}}\label{r_red}\end{equation}

\noindent where $\left\langle R^{2}\left(\beta Z,I\right)\right\rangle ^{1/2}$
is the root mean square (hereafter rms) nuclear charge radius. The
distribution of these reduced radii (normalized by a constant $r_{0}$,
as explained in the further part), based on experimental data \cite{Angeli_04}
for $N=574$ nuclei with $A$>100 is presented in fig. \ref{r1.epsi}
as a dashed line and exhibits variations about of 3\%.%

\begin{figure}[h]
\includegraphics[%
  width=8.5cm]{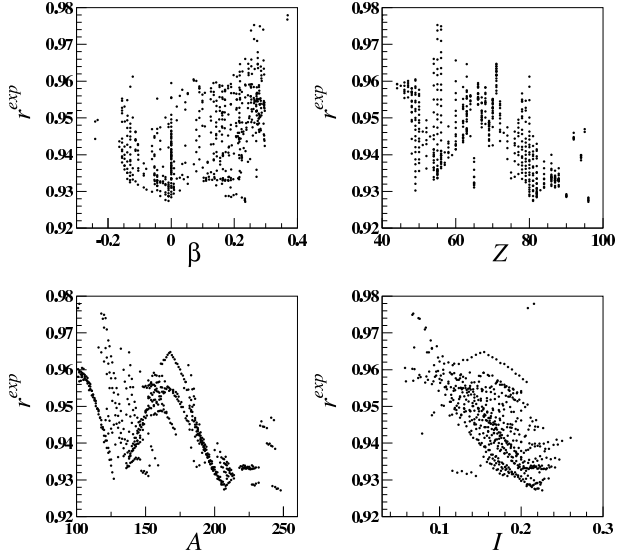}

\caption{\label{r2.epsi} Correlation between reduced radii and respectively
$\beta$, $Z$, $A$ and $I$. }
\end{figure}

\noindent Correlations between radius $r$ and $\beta$,$Z$, $A$
and $I$ parameters are presented on fig. (\ref{r2.epsi}).%

Looking at this figure, we cannot see clearly any systematical behavior.
This fact is related to confounded dependence on all parameters. Main
result of this letter is the procedure allowing us to unmask this
dependence and unravel generic correlations.

\noindent Since deviations in the rms distribution are small, it is
natural to adopt the form of an expansion, using simple parametric
form\begin{equation}
r=r_{0}\left(1+\alpha_{\beta}\cdot\beta^{2}\right)\left(1+\alpha_{Z}\cdot\left(Z-50\right)\right)\left(1+\alpha_{I}\cdot I\right)\label{r_fit}\end{equation}
 in which $\alpha_{\beta}$, $\alpha_{Z}$, $\alpha_{I}$ and $r_{0}$
are free coefficients. The dependence on $\beta^{2}$ results from
the dependence of rms radius on $\beta^{2}$ for quadruply deformed
shape. The form of the dependence on $I$ is justified by the results
obtained in the recent papers \cite{Sosin_06} where an important
observation was made, that for heavy nuclei, their nuclear rms charge
radii exhibit a systematic variation as a function of neutron-proton
asymmetry.

We can find coefficients $\alpha_{\beta}$, $\alpha_{Z}$, $\alpha_{I}$
and $r_{0}$ by considering distribution of value $\tilde{r}$ defined
as \begin{equation}
\tilde{r}=\frac{r^{exp}}{r_{0}\left(1+\alpha_{\beta}\cdot\beta^{2}\right)\left(1+\alpha_{Z}\cdot\left(Z-50\right)\right)\left(1+\alpha_{I}\cdot I\right)}\label{r_tyl}\end{equation}
 where the $r^{exp}$ is the experimental value of the reduced nuclear
charged radius. Now the distribution $F_{\alpha_{\beta}\alpha_{Z}\alpha_{I}r_{0}}\left(\tilde{r}\right)$
can be treated as a function of $\alpha_{\beta}$, $\alpha_{Z}$,
$\alpha_{I}$ and $r_{0}$. If the parametrization (\ref{r_fit})
is correct, the distribution of $\tilde{r}$ should tend to the Dirac
delta function $\delta(\tilde{r}-1)$ for some specific values of
$\alpha_{\beta}$, $\alpha_{Z}$, $\alpha_{I}$ and $r_{0}$. By minimizing
the variance of the distribution $F_{\alpha_{\beta}\alpha_{Z}\alpha_{I}r_{0}}\left(\tilde{r}\right)$
under the condition that its average value is equal to 1 the following
set of parameters was obtained:\begin{equation}
\alpha_{\beta}=0.225\;\;\alpha_{Z}=-0.00021\;\;\alpha_{I}=0.16\;\; r_{0}=0.9681\,\mathrm{fm}\label{fit}\end{equation}

To verify the equation (\ref{r_fit}), we plot on fig\emph{.} \ref{r2.epsi}
respective correlation and curves obtained from fit (\ref{r_fit}).
We notice the systematic and consistent behavior for almost all charge
radii, confirming the soundness of chosen parametrization.%
\begin{figure}[h]
\includegraphics[%
  width=8.5cm]{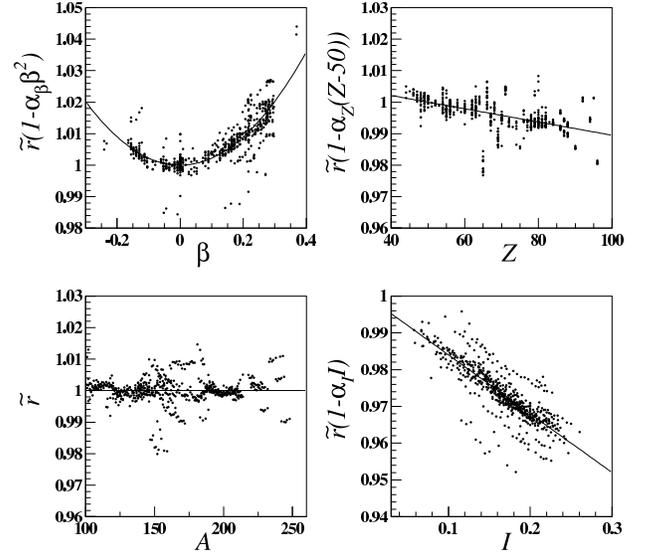}

\caption{\label{r3.epsi} Correlation between reduced radii with {}``switch-off''
dependence from value other than correlated one, and correlated values
are respectively $\beta$, $Z$, $A$ and $I$. Solid lines represents
for consider correlation fit given by (\ref{r_fit})}
\end{figure}

As we can see, for the majority of considered nuclei the chosen form
of parametrization works quite well. However, there are still some
groups of points on the plot which do not match the major trends.

To see the structure of these particular data better, for correlation
$\tilde{r}\left(1+\alpha_{I}\cdot I\right)\: vs.\: I$ we separate
this group by dashed lines given respectively by equation $0.988+\alpha_{I}\cdot I$
, $0.995+\alpha_{I}\cdot I$ and $1.005+\alpha_{I}\cdot I$. The points
located on the main branch are marked by green color whereas the points
above line $1.005+\alpha_{I}\cdot I$ and between lines $0.988+\alpha_{I}\cdot I$
and $0.995+\alpha_{I}\cdot I$ are marked by colors blue and yellow,
respectively. Eleven terbium points which are located below the line
$0.988+\alpha_{I}\cdot I$ were marked separately by red color. As
one can see green points match perfectly all correlations announced
by us. Other points follow also similar trends, but he corresponding
curves are slightly shifted comparing to the trends exhibited by green
points. Two reasons can be responsible for such behavior. First, the
systematics of the discrepancies can be associated with experimental
and systematical errors. Second, the effect may be the consequence
of some special properties of these nuclei. We should point out that
radii of terbium isotops (red points)  are measured within total (experimental
+ systematical) error about 3\% what is quite a large uncertainty
whereas the relative error of measured terbium radii with respect
to the radius of the reference isotope is much smaller and reads 0.04\%
\cite{Angeli_04}. It is therefore possible that that observed radii
deviations can origin from the total experimental error.

\begin{figure}[h]
\includegraphics[%
  width=8.5cm]{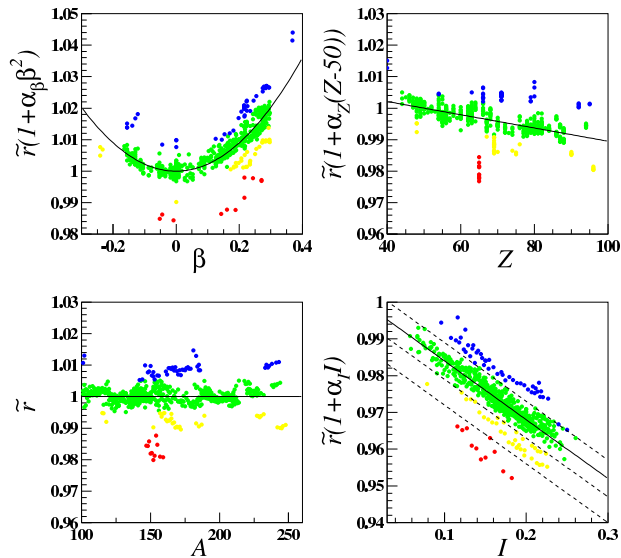}

\caption{\label{r4.epsi} Further analysis of correlations as presented in
Fig 3. Color of the points is determined by positions on correlations
between the isospin $I$ and $\tilde{r}\left(1+\alpha_{I}\cdot I\right)$.
The points from the main branch are marked by green color and are
located between dashed lines with respect to the equation $0.995+\alpha_{I}\cdot I$
and $1.005+\alpha_{I}\cdot I$. Points located above line $1.005+\alpha_{I}\cdot I$
are plotted in blue color whereas points below line $0.995+\alpha_{I}\cdot I$
and above line $0.988+\alpha_{I}\cdot I$ are plotted in yellow color.
Eleven points, located below line $0.988+\alpha_{I}\cdot I$ corresponding
to the terbium isotopes are plotted in red color.}
\end{figure}

We summarize our conclusions: 

\begin{enumerate}
\item Correlation observed by us proves that the functional form of radius
parametrization is correct for majority of consider nuclei. Ratio
of experimental to theoretical value for radii is presented on fig.
\ref{r1.epsi} as a solid line. The variation of this distribution
is equal to 0.0044, i.e. it corresponds to the value of experimental
errors.
\item The correlation between radius and the coefficient of deformation
clearly fits a parabolic shape and shows that in nuclear ground state
the dominating deformation is the quadrupole one. However it should
be stresses that value of $\alpha_{\beta}=0.225$ is more than two
times higher from the value expected for uniform quadruply deformed
matter. This is an interesting result which may point at strong correlations
between non-uniformity of the density of protons and quadrupole deformation.
Correlation between radii and isospin asymmetry is in agreement with
the earlier observations {[}submitted to \cite{Sosin_06} .
\item Effects of the surface can be easily taken into account, by adding
another functional parametric dependence with additional parameter
multiplying the $A^{-1/3}$ term. We have found that these effects
are negligible. This observation is consistent with our choice of
considering only heavy nuclei, where surface effects are not expected
to play the major role.
\item Correlation between radii $r$ and $Z$ shows, that at value of deformation
and neutron-proton asymmetry fixed, protons become on average closer,
as the number of protons increases. This observation probably suggests
the possibility of the creation of the neutron skin. This correlation
accompanied by the correlation between radii and the isospin can impose
interesting constraints on the form of the nuclear equation of state.
This possibility would be addressed in the forthcoming publication.
\item Taken into account the systematics of deviations for the terbium isotopes,
it would be interesting to obtained and compare experimental data
corresponding to smaller errors. In the \emph{}future the shift from
systematical behavior \emph{}also \emph{}for isotopes represented
by blue \emph{}and yellow points should be considered and explained.
\end{enumerate}
\begin{acknowledgments}
The author is indebted to E. Nowak, M. A. Nowak and P. Pawlowski for
careful reading of manuscript and helpful discussions. This work was
supported the Polish Ministry of Science, research project no. PB
1 P03B 020 30.
\end{acknowledgments}

\end{document}